\documentstyle[12pt]{article}

\input amssym.def       
\input amssym.tex       

\def\goth{\frak}          
\def\double{\Bbb}         
\def\ccal{\cal}           

\def\cc{{\double C}}     
       
\def\rr{{\double R}}     

\def\aa{{\cal A}}
\def\dd{{\cal D}}
\def\gg{{\goth g}}        
\def\hh{{\cal H}}
\def\hhh{{{\double H}}}   
\def\mm{{{\ccal M}}}

\def\t{\mathrm{Tr}} 
\def\tp{\mathrm{Tr_p}}
\def\lb{\left[} 
\def\rb{\right]}
\def\lp{\left(} 
\def\rp{\right)}
\def\la{\left\{} 
\def\ra{\right\}}

\def\ov{\overline}
\def\ot{\otimes}
\def\op{\oplus}

\def\bb{\begin{eqnarray}}
\def\ee{\end{eqnarray}}
\def\eee{\end{eqnarray}}
\def\pp{\pmatrix}

\begin{document}

\hsize 17truecm
\vsize 24truecm
\font\twelve=cmbx10 at 13pt
\font\eightrm=cmr8
\baselineskip 18pt

\begin{titlepage}

\centerline{\twelve CENTRE DE PHYSIQUE THEORIQUE}
\centerline{\twelve CNRS - Luminy, Case 907}
\centerline{\twelve 13288 Marseille Cedex 9}
\vskip 4truecm
\begin{center}
{\bf Thomas KRAJEWSKI}
\footnote{ and Universit\'e de Provence and Ecole Normale Superieure de Lyon, tkrajews@cpt.univ-mrs.fr} \\
\bf Igor PRIS
\footnote{ and Universit\'e de Provence, pris@cpt.univ-mrs.fr \\
and Institute of Physics, Belarus, pris@bas11.basnet.minsk.by } \\
\end{center}
\vskip 2truecm

\centerline{{\bf\large  \sc Towards a $Z'$ Gauge Boson in Noncommutative Geometry  }}
\bigskip
\vskip 2truecm
\leftskip=1cm
\rightskip=1cm
\centerline{\bf Abstract} 

\medskip
We study all possible $U(1)$-extensions of the standard model (SM) in the framework of noncommutative geometry (NCG) with the algebra $\hhh\op\cc\op\cc\op M_3(\cc)$. Comparison to experimental data about the mass of a hypothetical $Z'$ gauge boson leads to the necessity of introducing at least one new family of heavy fermions.
 
\vskip 1truecm
PACS-92: 11.15 Gauge field theories\\ 
\indent
MSC-91: 81E13 Yang-Mills and other gauge theories\\ 
\vskip 1truecm

\noindent June 1996\\
\vskip 0.2truecm
\noindent
CPT-96/P.3360\\
hep-th/9607005
\end{titlepage}

According to recent works (see for example \cite{exp1}-\cite{exp2}), experimental data provide us with some deviations from the SM that can be explained by introducing a new $U(1)$ gauge  boson, called $Z'$ gauge boson, with mass bigger than 1 Tev. Therefore, the topic of the $U(1)$ extensions of the SM arises as a natural question in particle physics. There are several answers to this question (for example for superstrings derived $Z'$ see \cite{strings}). In this paper we use the approach of NCG, developped by A. Connes (see \cite{Co1}-\cite{Co4}, \cite{KM}), to construct some $U(1)$ extensions of the SM without adding new fermions. In the framework of NCG a class of Yang-Mills-Higgs (YMH) theories can be entirely geometrized \cite{IS}: the gauge and Higgs fields are obtained as the two components of a single 1-form on a noncommutative space-time. This noncommutative space-time is taken to be the product of ordinary space-time by a discrete noncommutative (the "internal space") that reminds us of the internal degrees of freedom of the YMH theory. All data concerning the geometry of this noncommutative space-time are encoded in a {\it spectral triple} $(\aa_t,\hh_t,D_t)$ where $\aa_t$ is an associative algebra represented on a Hilbert space $\hh_t$ and $\dd_t$ is a hermitian operator of $\hh_t$ that contains the fermionic sector of the theory and especially the masses and mixing terms for fermions. This is the starting point of the construction of the differential algebra (a noncommutative generalization of the de Rham complex), and we retrieve as an output the whole action of the YMH theory. In particular, the Higgs potential is expressed as a function of the fermionic mass and mixing terms and then, we expect the masses of the gauge bosons to be related to these masses and mixing terms. This is the keystone of this work, and our purpose is to answer the following question: can we incorporate a heavy gauge boson into the SM in the framework of NCG without changing its fermionic sector?\\
In NCG, the gauge group $G$ is obtained as the group of unitary elements of an involutive and associative algebra $\aa$ after application of a unimodularity condition (this means that a $U(1)$ factor is eliminated in the group of unitary elements of the algebra $\aa$). For the SM the algebra $\aa$ is $\hhh\op\cc\op M_{3}(\cc)$ where $\hhh$ stands for the field of quaternions and $M_{3}(\cc)$ for the algebra of $3\times 3$ complex matrices. There are only two ways to get an additional $U(1)$ gauge boson: either we can replace the quaternions by the two by two complex matrices (this variant has been considered in \cite{PS}) or we can add to the algebra $\hhh\op\cc\op M_{3}(\cc)$ of the SM one more factor~$\cc$.\\
Here we explore the second possibility. To do so we first have to select all the representations of the algebra $\hhh\op\cc\op\cc\op M_{3}(\cc)$
(we do not change the other inputs of NCG) that satisfy both geometrical and physical constraints. Once we have all acceptable representations, we construct the associated YMH theory and compute an upper bound of the masses of the gauge bosons.\\
Let us begin by giving a brief review of the construction of a YMH theory in NCG.
\section{$U(1)$ extensions of the SM and spectral triples}
\subsection{NCG and YMH theory}
NCG provides us with a tool to build some YMH theories. In this framework, we usually consider the product of the ordinary geometry of a four dimensional euclidean and compact manifold $\mm$ with a noncommutative internal space. All the data of this geometry can be encoded in a spectral triple $(\aa_t,\hh_t,D_t)$ which is a product of two spectral triples corresponding to these two spaces:
\bb
\aa_t&\equiv&C^{\infty}(\mm)\ot\aa,\nonumber\\
\hh_t&\equiv&L^{2}(S,\mm)\ot\hh,\nonumber\\
D_t&\equiv&i\gamma^{\mu}\partial_{\mu}\ot I+\gamma^{5}\ot D,\nonumber
\eee
where $C^{\infty}(\mm)$ is the commutative algebra of complex valued smooth functions on $\mm$ represented on the Hilbert space of square integrable Dirac spinors  $L^{2}(S,\mm)$ by usual multiplication and $\gamma^{\mu}$ are the euclidiean and hermitian Dirac matrices. $(\aa,\hh,D)$ is a spectral triple for the internal space that consists of an involutive and faithful representation $\rho$ of an involutive finite dimensional algebra $\aa$ in a finite dimensional Hilbert space $\hh$ of all fermions and antifermions involved in the theory and a hermitian operator $D$ of $\hh$, called Dirac operator, that contains all fermion masses and mixing terms. To obtain a generalization of the ordinary Riemannian geometry on our noncommutative space time, the spectral triple $(\aa_t,\hh_t,\dd_t)$ has to satisfy some axioms involving two other operators (\cite{Co4}): the charge conjugation $J_t$ (that exchanges particles and antiparticles) and the chirality $\chi_t$ (that selects left-handed and right-handed particles). For our product geometry, these two operators are given by the tensor product of the usual charge conjugation and chirality on Dirac spinors by   the operators $J$ and $\chi$ of the finite dimensional Hilbert space $\hh$. Further, since we exclude Majorana particles, the spectral triple $(\aa,\hh,\dd)$ is assumed to be $S^{0}$-real \cite{Co3}. The most general $S^{0}$-real spectral triple $(\aa,\hh,D)$ can be written in the following manner:
\bb
&\rho=\pp{\rho_{L}&0&0&0\cr
0&\rho_{R}&0&0\cr
0&0&\rho_{L}^{c}&0\cr
0&0&0&\rho_{R}^{c}},\>\>\>\> 
D=\pp{0&M&0&0\cr
M^{*}&0&0&0\cr
0&0&0&\ov{M}\cr
0&0&\ov{M}^{*}&0}&\nonumber 
\eee
\bb
&\chi=\pp{-I&0&0&0\cr
0&I&0&0\cr
0&0&-I&0\cr
0&0&0&I},\>\>\>\>
J=\pp{0&0&I&0\cr
0&0&0&I\cr
I&0&0&0\cr
0&I&0&0}C,&
\eee
where the subscripts L and R refer to left-handed and right-handed fermions and the superscript c refers to antifermions, $M$ is the mass and mixing matrix of all fermions, $C$ is the complex conjugation and the $I$ denotes the identity matrix  with suitable dimension.\\
In this case, all the axioms of NCG reduce to {\it reality}, {\it Poincar\'e duality} and {\it orientability} axioms. We will give more details about these axioms in the next section.\\
Once we have a spectral triple $(\aa_t,\hh_t,D_t)$ we define a
differential algebra that we represent on the Hilbert space $H_t$ (see \cite{KPPW} for a general construction of the differential algebra in NCG). Roughly
speaking, this representation is obtained by replacing the exterior
derivative by commutators with $D_t$, and in particular, the gauge fields
and the Higgs field appear as an antihermitian 1-form on the
product space. Due to the special form of $D_t$, this 1-form splits
into the gauge fields and the Higgs field. Assuming implicit space time dependance, the latter can be written as
a purely noncommutative 1-form:
\bb
H&\equiv&\sum_{i}\rho(x_{i}^{0})\lb D,\rho(x_{i}^{1})\rb\>\>\>\> \mathrm{with}\; x_{i}^{0}\;\mathrm {and}\; x_{i}^{1}\in\aa,\label{Higgs}
\eee 
and its curvature $C$ is defined by:
\bb
C&\equiv&\delta H+H^{2}\label{curv},
\eee 
where $\delta$ is the derivation of our internal differential algebra.\\
Then, the preliminary Higgs potential is
\bb
V_{0}(H)\equiv Tr(zC^{2}).\label{prelpot}
\eee
In the previous formula, $z$ denotes a positive definite matrix that commutes with $\rho(\aa)$, $J\rho(\aa)J$ and $D$, such that
\bb
(\omega,\omega')&\equiv&\t(z\omega^{*}\omega')\label{scalprod}
\eee
is a scalar product between two forms $\omega$ and $\omega'$ of same degree (see \cite{Ka}-\cite{IKS}).
As a consequence, the Higgs field and the Higgs potential have a geometrical interpretation in NCG because the Higgs field can be considered as a gauge field on the internal space and the Higgs potential is the square norm of its curvature.\\
We end this section by giving the explicit form of the spectral triple of the SM with the notations introduced at the beginning of this section:
\bb
\rho_{L}(a)&\equiv&\mathrm{diag}\lp a\ot I_{3N},\,a\ot I_{N}\rp\nonumber,\\
\rho_{R}(b)&\equiv&\mathrm{diag}\lp b\,I_{3N},\,\ov{b}\,I_{3N},\,\ov{b}\,I_{N}\rp\nonumber,\\
\rho_{L}^{c}(b,c)&\equiv&\mathrm{diag}\lp I_{2N
}\ot c,\,\ov{b}\,I_{2N}\rp\nonumber,\\
\rho_{R}^{c}(b,c)&\equiv&\mathrm{diag}\lp
I_{2N}\ot c,\,\ov{b}\,I_{N}\rp\nonumber,
\eee
where $(a,b,c)\in\hhh\op\cc\op M_{3}(\cc)$ and N stands for the number of families. The mass matrix $M$ is given by:
\bb
M&\equiv&\pp{\pp{M_{u}\ot I_{3}&0\cr 0&M_{d}\ot I_{3}}&0\cr
0&\pp{0\cr M_{e}}},\nonumber
\eee
where
\bb
M_{u}\equiv\mathrm{diag}\lp m_{u},m_{c},m_{t}\rp, \>
M_{d}\equiv V_{CKM}\,\mathrm{diag}\lp m_{d},m_{s},m_{b}\rp, \>
M_{e}\equiv\mathrm{diag}\lp m_{e},m_{\mu},m_{\tau}\rp,\nonumber
\eee
 $m_{p}$ stands for the mass of particle $p$ and $V_{CKM}$ is the Cabibbo-Kobayashi-Maskawa mixing matrix. The chirality $\chi$ and the charge conjugation $J$ are given by
\bb
\chi&\equiv&\mathrm{diag}\lp -I_{8N},I_{7N},-I_{8N},I_{7N}\rp, \nonumber
\eee
\bb
J&\equiv&\pp{0&I_{15N}\cr I_{15N}&0}C,\nonumber
\eee
with $C$ denoting the complex conjugation. Finally, the noncommutative gauge coupling $z$ has the form
\bb
&z\;\equiv\;\mbox{diag}\lp x/3I_{3N},\;I_{2}\ot y,\;
x/3I_{3N},\;y,\; 
\tilde x/3I_{3N},\;I_{2}\ot \tilde y,\;
\tilde x/3I_{3N},\;\tilde y\rp,&\label{coup}
\eee
where $x$ and $\tilde x$ are strictly positive numbers and $y$ and $\tilde y$ positive definite diagonal matrices of size $N\times N$.\\
In the next section, we use the axioms of NCG to select all the representations of $\hhh\op\cc\op\cc\op M_{3}(\cc)$ that provide us with $U(1)$ extensions of the SM.
\subsection{Geometrical constraints}
The choice of a spectral triple $(\aa_t,\hh_t,D_t)$ is not arbitrary
and should satisfy the set of  NCG axioms \cite{Co4}. As already pointed, the axioms of NCG reduce in our case to reality, Poincar\'e duality and orientability axioms since in the case of a product of usual Riemannian geometry with a $S^{0}$-real spectral triple all other axioms are obviously satisfied. The reality is equivalent to the following two
relations: 
\begin{equation}
\lb\rho(x),J\rho(x')J\rb=0, \>\>\>\>
\lb\lb D,\rho(x)\rb,J\rho(x')J\rb=0, \>\>\>\> x,x'\in\aa.   \label{Poinc}       
\end{equation}           
The Poincar\'e duality means the non-degeneracy  
of the intersection form which is given 
by the matrix $\cap$:
\begin{equation}
\cap_{ij}=\t(\chi\rho(p_{i})J\rho(p_{j})J),   \label{int}  
\end{equation}
where   $p_{i}$ are the minimal hermitian projections of our algebra.\\
The  orientability axiom means that the chirality can be written in the following manner 
\begin{equation}
\chi=\sum_{i}\rho(a_{i})J\rho(b_{i})J \>\>\>\>\>\> a_i,b_i\in\aa. \label{orient}
\end{equation}
Now, we modify the spectral triple of the SM in order to obtain a $U(1)$ extension. To do so, we leave the Hilbert space $\hh$ and the operators $J$, $\chi$ and $D$ unchanged and we modify the representation of the algebra of the SM to obtain a representation 
$\rho$ of the algebra $\hhh\op\cc\op\cc\op M_{3}(\cc)$ that takes $(a,b,b',c)\in\hhh\op\cc\op\cc\op M_{3}(\cc)$ to $\rho(a,b,b',c)$, without changing the weak and strong sectors. Then, (\ref{Poinc})
show that $\rho_{L}$, $\rho_{L}^{c}$ and $\rho_{R}^{c}$ remain the same and $\rho_{R}$ can be can be written:
\bb
\rho_{R}(b,b')&=&\mathrm{diag}\lp\alpha\,I_{3N},\,\beta\,I_{3N},\,\gamma\, I_{N}\rp,\nonumber
\eee
with $\alpha,\beta,\gamma\in\la b,\ov{b},b',\ov{b}'\ra$.\\
It is easy to check that $\chi=\rho(-I_{2},1,1,I_{3})J\rho(-I_{2},1,1,I_{3})J,$
so that axiom (\ref{orient}) is fulfilled.\\
Further, the determinant  of the intersection form  (\ref{int}) is equal to zero for some particular distribution in the representation $\rho$ of the two $\cc$ summands of the algebra. Once we have eliminated these representations, the axioms of NCG leave $4^{3}-3\times 2^{3}=40$ possibilities that are listed in the following table:
\begin{center}
\begin{tabular}{|c|c|c|c|} 
\hline
$\alpha$&$\beta$&$\gamma$&$det\cap$\\	
\hline			 
$b,\;\ov{b}$&$b,\;\ov{b}$&$b,\;\ov{b}$&$ =0$\Big.\\
\hline			 
$b,\;\ov{b}$&$b,\;\ov{b}$&$b',\;\ov{b}'$&$\ne 0$\Big.\\
\hline			 
$b,\;\ov{b}$&$b',\;\ov{b}'$&$b,\;\ov{b}$&$\ne 0$\Big.\\
\hline			 
$b',\;\ov{b}'$&$b,\;\ov{b}$&$b,\;\ov{b}$&$\ne 0$\Big.\\
\hline			 
$b,\;\ov{b}$&$b',\;\ov{b'}$&$b',\;\ov{b}'$&$=0$\Big.\\
\hline			 
$b',\;\ov{b}'$&$b',\;\ov{b}'$&$b,\;\ov{b}$&$\ne 0$\Big.\\
\hline			 
$b',\;\ov{b}'$&$b,\;\ov{b}$&$b',\;\ov{b}'$&$=0$\Big.\\
\hline			 
$b',\;\ov{b}'$&$b',\;\ov{b}'$&$b',\;\ov{b}'$&$\ne 0$\Big.\\
\hline
\end{tabular}
\end{center}
Note that since the intersection form involves hermitian projections, it does not distinguish a representation from its complex conjugate.\\
Although these 40 spectral triples satisfy all geometrical axioms, we analyse in the next section the physical constraints: the anomaly cancellation and the electric charge. 
\newpage
\subsection{Physical constraints}
\subsubsection{Unimodularity}
In NCG, the gauge group $G$ is obtained from the group of unitary elements of the algebra $\aa$ after application of a unimodularity condition that cancels a $U(1)$ factor \cite{Co4}-\cite{MGV}. In our case, the unimodularity condition is:
\begin{equation}
\tp\lb\rho(a,b,b',c)+J\rho(a,b,b',c)J\rb=0,
\label{unim}
\end{equation}
where $i(a,b,b',c)\in su(2)\op i\rr\op i\rr\op u(3)$ and $\mathrm{Tr}_{\mathrm p}$ is the trace on $\hh$ restricted to particle space.\\ 
The unimodularity condition allows us to reexpress the $u(1)$ Lie algebra coming from the decomposition $u(3)=u(1)\op su(3)$ of the Lie algebra of the unitaries of $M_3(\cc)$ in terms of the two $U(1)$ arising from the two $\cc$ summands of $\aa$. From now on we note $\gg=su(2)\op i\rr\op i\rr\op su(3)$ the Lie algebra of the group of the unitary elements of $\aa$ (after application of the unimodularity condition) whose representation is $\tilde{\rho}=\mathrm{diag}\lp\tilde{\rho}_{L},\tilde{\rho_{R}},\tilde{\rho}_{L}^{c},\tilde{\rho}_{R}^{c}\rp$ on Hilbert space of particles and antiparticles with:
\bb
\tilde{\rho}_{L}(a)&=&\mathrm{diag}\lp a\ot I_{3N},a\ot I_{N}\rp,\nonumber\\
\tilde{\rho}_{R}(b,b')&=&\mathrm{diag}\lp
\lp y_{u}b+y_{u}^{'}b'\rp\,I_{3N},\,
\lp y_{d}b+y_{d}^{'}b'\rp\,I_{3N},\,
\lp y_{e}b+y_{e}^{'}b'\rp\,I_{N},\,
\rp,\nonumber\\
\tilde{\rho}_{L}^{c}(b,b',c)&=&\mathrm{diag}\lp
I_{2N}\ot\lp c+ubI_{3}+u'b'I_{3}\rp,\,
-bI_{2N}
\rp,\nonumber\\
\tilde{\rho}_{R}^{c}(b,b',c)&=&\mathrm{diag}\lp
I_{2N}\ot\lp c+ubI_{3}+u'b'I_{3}\rp,\,
-bI_{N}\rp,\nonumber
\eee
where  $(a,b,b',c)\in i\gg$, $y,y'\in\la 1,0,+1\ra$ (the generalized hypercharges) and $u$ and $u'$ are rational linear combinations of the generalized hypercharges.

\subsubsection{Anomalies}

We note here that in the NCG framework, for the SM \cite{AGM}, the 
unimodularity  condition  (\ref{unim}) is equivalent to the condition of cancellation of gauge anomalies. In general, this condition is
\begin{equation}
\tp\lb\chi\lp\tilde\rho (x)+J\tilde\rho(x)J\rp^3\rb=0  \>\>\>\>\>\> \mbox{for all} \>\>\> x\in i\gg. 
\label{gano}     \end{equation}
Unfortunately, in our case, the gauge anomaly cancellation (\ref{gano}) is not equivalent to the unimodularity condition (\ref{unim}). Moreover, from (\ref{gano}) it is easy to see
that all $U(1)$ extension of the SM are always anomalous. As our group consists of 4 factors, asking for anomaly cancellation is too restrictive and cannot be met  without
adding new fermions even in the 
much larger, YMH framework,
which we adopt in this section. (As it was shown in \cite{IS}
every model arising from a spectral triple is a particular case
of the  general YMH  model.)   
 
 Restricting  to one fermion family and 
leaving the $SU(2)$ and $SU(3)$ content of the standard model 
fermion representation untouched, our starting point is the
fermion representation symmetrized with
respect to charge conjugation and restricted
to particle space with basis  $(u,d)_L$, $(\nu, e)_L$, $u_R, d_R, e_R$:
\bb
\tilde\rho(a,b,b',c)+J\tilde\rho(a,b,b',c)J=
\mbox{diag} \pp{a\ot 1_3+1_2\ot c+y_1 b 1_6+y_1'b'1_6\cr
a+y_2 b 1_2+y_2'b'1_2\cr
c+y_3b1_3+y_3'b'1_3\cr
c+y_4b1_3+y_4'b'1_3\cr
y_5b+y_5'b'}
\eee 
for all $(a,b,b',c)\in i\gg.$
The 
"hypercharges" $y_i$ and $y_i'$ are arbitrary real numbers.
In the noncommutative framework, they can only take the 
values -1,0,1 with the constraints arising from NCG.

In terms of the 
hypercharges, the  condition  of vanishing gauge anomalies (\ref{gano}) is  equivalent to
\bb
6y_1^3+2y_2^3-3y_3^3-3y_4^3-y_5^3=0\label{8} \\  
3y_1+y_2=0 \label{9}                          \\  
6y_1-3y_3-3y_4=0\label{10}                       \\
6y_1+2y_2-3y_3-3y_4=0\label{11}                   \\
6y_1^2y_1'+2y_2^2y_2'-3y_3^2y_3'-3y_4^2y_4'-y_5^2y_5'=0\label{13}  
\eee
and the same equations with $y$ and $y'$ interchanged.

Let us first solve the SM case  (\ref{9})-(\ref{11}), $y_i'=0$.
There are two families of solutions:
a two-parameter family with parameters
\bb
y_5&<&0 \> \>\>\>\>\>\mbox{and}\>\>\>\>\>\> y_1\in \left(0,-3^{-1}2^{-\frac{1}{3}}y_5\right)\nonumber
\eee
given by
\begin{equation}
y_2=-3y_1,\;
y_3=2y_1-y_4,\;
y_4=y_1\left(1\pm\sqrt{3}\sqrt{\frac{1}{2}\left(-y_5/3y_1\right
)^3-1}\right)\nonumber
\end{equation}
and a one-parameter family with parameter $y_3$
given by
\bb
y_1=0,\>\>\> y_2=0,\>\>\> y_4=-y_3,\>\>\> y_5=0.\nonumber
\eee

The SM belongs to the first family with
$y_{5}=-1$ and $y_1=1/6$. The second family is
the bizarre (hadrophilic) solution \cite{MRW}.  In parenthesis, we remark that
adding the condition of vanishing 
gravitational anomalies \cite{AW}: 
\bb
\tp\lb\chi\lp\tilde\rho(x)+J\tilde\rho(x)J\rp\rb=0  \>\>\>\> \>\>\mbox{for all} \>\>\> x\in i\gg
\nonumber
\eee
or equivalently,
\bb
6y_1+2y_2-3y_3-3y_4-y_5=0\label{12}             
\eee
reduces the first family precisely to the SM,
and leaves the second family one parameter.

(\ref{8})-(\ref{11}) and the same equations involving $y'$ say that the primed hypercharges
fall on similar two and one dimensional
submanifolds in the general hypercharge manifold.
Finally, (\ref{13}) and the same equation with $y$ and $y'$ interchanged say that 
any linear combination 
$my_i+ny_i'$ of solutions $y_i$  and $y_i'$
must be a solution implying that the $Z$ and
$Z'$ must have identical couplings to fermions. 
\subsubsection{Electric charge}
In the previous section, we have left untouched the weak and strong sectors of the SM. Since we want our model to be an extension of the SM, we have to take care of the electric charge. In NCG, the  fermionic action for particles is: $<\psi,\lp D+\tilde\rho(x)+J\tilde\rho(x)J\rp\psi>$ where $\psi$  is a vector in the particle space and $ix\in\gg$. Since we want to retrieve the electric charge $Q$, we have to find an element $x\in i\gg$ such that $Q=\tilde\rho(x)+J\tilde\rho(x)J$. This leads to a linear system with unknown $x$ that is compatible only for some values of the generalized hypercharges and thus  select some of the 24 spectral triples considered in the previous section which are listed and classified into four types of spectral triples. We give in the following table the generalized hypercharges, the coefficients $u$ and $u'$ and the type of each of the 12 representations that satisfy both  geometrical and physical axioms:
\bigskip
\begin{center}
\begin{tabular}{|c|c|c|c|c|c|c|c|c|}
\hline
$y_{u}$&$y_{u}'$&$y_{d}$&$y_{d}'$&$y_{e}$&$y_{e}'$&$u$&$u'$&type\\
\hline
$ 0 $&$ 1 $&$ 0 $&$ -1 $&$ 0 $&$ -1 $&$ 1/4 $&$ 1/12 $&$ 1 $\\
\hline
$ 0  $&$ -1 $&$ 0 $&$ 1 $&$ 0 $&$ 1 $&$ 1/4 $&$ -1/12 $&$ 1 $\\
\hline
$ 1 $&$ 0 $&$ -1 $&$ 0 $&$ 0 $&$ 1 $&$ 1/4 $&$ -1/12 $&$ 2 $\\
\hline
$ 0 $&$ 1 $&$ 0 $&$ -1 $&$ -1 $&$ 0 $&$ 1/3 $&$ 0 $&$ 2 $\\
\hline
$ 1 $&$ 0 $&$ -1 $&$ 0 $&$ 0 $&$ -1 $&$ 1/3 $&$ 0 $&$ 2 $\\
\hline
$ 1 $&$ 0 $&$ -1 $&$ 0 $&$ 0 $&$ -1 $&$ 1/4 $&$ 1/12 $&$ 2 $\\
\hline
$ 0 $&$ -1 $&$ -1 $&$ 0 $&$ 0 $&$ 1 $&$ 1/2 $&$ 1/6 $&$ 3 $\\
\hline
$ 1 $&$ 0 $&$ 0 $&$ -1 $&$ -1 $&$ 0 $&$ 1/12 $&$ 1/4 $&$ 3$\\
\hline
$ 1 $&$ 0 $&$ 0 $&$ 1 $&$ -1 $&$ 0 $&$ 1/12 $&$ -1/4 $&$ 3 $\\
\hline
$ 0 $&$ 1 $&$ -1 $&$ 0 $&$ 0 $&$ -1 $&$ 1/2 $&$ -1/6 $&$ 3 $\\
\hline
$ 1 $&$ 0 $&$ 0 $&$ 1 $&$ 0 $&$ 1 $&$ 0 $&$ -1/3 $&$4$\\
\hline
$ 0 $&$ 1 $&$ -1 $&$ 0 $&$ -1 $&$ 0 $&$ 7/12 $&$ -1/4 $&$ 4 $\\
\hline
$ 0 $&$ -1 $&$ -1 $&$ 0 $&$ -1 $&$ 0 $&$ 7/12 $&$ 1/4 $&$ 4$\\
\hline
$ 1 $&$ 0 $&$ 0 $&$ -1 $&$0 $&$-1$&$ 0 $&$ 1/3 $&$ 4$\\
\hline
\end{tabular}
\end{center}
The interest of this spectral triples classification appears when we compute the Higgs; all spectral triples of the same type give rise to the same Higgs field. Furthermore, it is particular case of a general classification \cite{Kr}.
\section{Spontaneous symmetry breaking and $Z'$ mass}

In this section, we evaluate the mass of the extra $U(1)$ gauge boson. We first determine the Higgs field and the minimum of its potential and then, use this minimum to compute the mass term for the gauge bosons after spontaneous symmetry breaking.

\subsection{The Higgs field}
Remember that in NCG the Higgs field is an antihermitian 1-form given by (\ref{Higgs}):
\bb
H=\sum_{i}\rho(x_{i}^{0})\lb D,\rho(x_{i}^{1})\rb\;\mbox{with}\;x_{i}^{0}\;\mbox{and}\;x_{i}^{1}\in\aa.\nonumber
\eee
According to Poincar\'e duality, the commutator $\lb D,\rho(\aa)\rb$ vanishes on antiparticle space ($\rho^c$ is vector like) and from now on, we restrict all matrices to particle space. To recover the genuine Higgs field of a YMH theory, we perform the change of variable $\Phi=H-iD$. Notice that the field $\Phi$ is an antihermitian 1-form that transforms homogeneously under a gauge transformation, that is, a gauge transformation parameterized by $u\in G$ maps $\Phi$ to $\rho(u^*)\Phi\rho(u)$. Using the new variable, the curvature $C$ defined by (\ref{curv}) is:
\begin{equation}
C=\Phi^{2}+D^{2}+\theta,
\end{equation}
where $\theta$ is an element of the junk in degree two $J^{2}$ determined by the condition that the curvature is orthogonal to $J^{2}$ in the sense of the scalar product (\ref{scalprod}) defined by the noncommutative gauge coupling $z$. Since $z$ has to commute with $\rho(\aa)$, $J\rho(\aa)J$ and $D$, we can easily prove that it is given by the parametrization used for the gauge coupling of the SM as written in equation (\ref{coup}).\\
The junk appears in NCG when we try to represent the formal differential algebra on the Hilbert space by replacing the derivative by a commutator with the Dirac operator. To obtain a representation of the differential structure, we have to divide by a two-sided ideal called junk; degree two refers to 2-forms. For our purpose, it is sufficient to know that a generic element $\theta$ of $J^{2}$ is given by
\bb
\theta&=&\sum \lb D,\rho(x_{i}^{0})\rb\lb D,\rho(x_{i}^{1})\rb, \nonumber
\eee 
with the condition 
\bb
\sum \rho(x_{i}^{0})\lb D,\rho(x_{i}^{1})\rb&=&0,\nonumber
\eee
where $x_{i}^{0},x_{i}^{1}\in\aa$.
It is not difficult to prove that in all cases, the junk in degree two is
\bb
J^{2}&=&\la\pp{ih\ot\Delta_{q}&0&0&0\cr
0&ih\ot\Delta_{l}&0&0\cr
0&0&0&0\cr
0&0&0&0},\;h\in\hhh\:\ra,  \nonumber
\eee
where
$  
\Delta_{q}=\frac{1}{2}(M_{u}M_{u}^{*}-M_{d}M_{d}^{*}), \>\>\>
\Delta_{l}=-\frac{1}{2}M_{e}M_{e}^{*}.
$
Then, it is easy to compute the Higgs field $\Phi$ and its curvature $C$, and we give the results in our four types of models, using the following notations:
\bb
\Phi\;=\;i\pp{0&0&\Phi_{q}&0\cr
0&0&0&\Phi_{l}\cr
\Phi_{q}^{*}&0&0&0\cr
0&\Phi_{l}^{*}&0&0}&,&
C\;=\;\pp{C_{q}^{L}&0&0&0\cr
0&C_{l}^{L}&0&0\cr
0&0&C_{q}^{R}&0\cr
0&0&0&C_{l}^{R}}  \nonumber
\eee
\begin{itemize}
\item
Type 1 models, one scalar doublet: $\phi$ 
\bb
\Phi_{q}\;=\;(\phi\ot I_{N}\ot I_{3})M_{q}
&,&\Phi_{l}\;=\;(\phi\ot I_{N})M_{l}  \nonumber\\
C_{q}^{L}\;=\;(1-\phi\phi^{\dag})\ot\Sigma_{q}&,&
C_{l}^{L}\;=\;(1-\phi\phi^{\dag})\ot\Sigma_{l}\nonumber\\
C_{q}^{R}\;=\;M_{q}^{\dag}(1-\phi\phi^{\dag})M_{q}&,&
C_{l}^{R}\;=\;M_{l}^{\dag}(1-\phi\phi^{\dag})M_{l}  \nonumber
\eee
\item
Type 2 models, two scalar doublets: $\phi_l$ and $\phi-q$
\bb
&\Phi_{q}\;=\;(\phi_{q}\ot I_{N}\ot I_{3})M_{q}\;,\;
\Phi_{l}\;=\;(\phi_{l}\ot I_{N})M_{l},&\nonumber\\
&C_{q}^{L}\;=\;(1-\phi_{q})\phi_{q}^{\dag}\ot\Sigma_{q}+
(\phi_{l}\sigma_{3}\phi_{l}^{\dag}-\phi_{q}\sigma_{3}\phi_{q}^{\dag})
\ot\Delta_{q}^{'},&\nonumber\\
&C_{l}^{L}\;=\;(1-\phi_{l})\phi_{l}^{\dag}\ot\Sigma_{l}+
(\phi_{q}\sigma_{3}\phi_{q}^{\dag}-\phi_{l}\sigma_{3}\phi_{l}^{\dag})
\ot\Delta_{l}^{'}&\nonumber\\
&C_{q}^{R}\;=\;M_{q}^{\dag}(1-\phi_{q}\phi_{q}^{\dag})M_{q}\;,\;
C_{l}^{R}\;=\;M_{l}^{\dag}(1-\phi_{l}\phi_{l}^{\dag})M_{l},&\nonumber
\eee
\item
Type 3 ($\epsilon=-1$) and Type 4 ($\epsilon=+1$), two scalar doublets: $\phi_l$ and $\phi_q$
\bb
&\Phi_{q}\;=\;((\phi_{1}+\phi_{2}\sigma_{3})\ot I_{N}\ot I_{3})M_{q}\;,\;
\Phi_{l}\;=\;((\phi_{1}+\epsilon\phi_{2}\sigma_{3})\ot I_{N})M_{l},&\nonumber\\
&C_{q}^{L}\;=\;(1-\phi_{1}\phi_{1}^{\dag}-\phi_{2}\phi_{2}^{\dag})\ot\Sigma_{q}
-(\phi_{1}\phi_{2}^{\dag}+\phi_{2}\phi_{1}^{\dag})\ot\Delta_{q}
-(\phi_{1}\sigma_{3}\phi_{2}^{\dag}+\phi_{2}\sigma_{3}\phi_{1}^{\dag})
\ot\Sigma_{q}^{'},&\nonumber\\
&C_{l}^{L}\;=\;(1-\phi_{1}\phi_{1}^{\dag}-\phi_{2}\phi_{2}^{\dag})\ot\Sigma_{l}
-\epsilon(\phi_{1}\phi_{2}^{\dag}+\phi_{2}\phi_{1}^{\dag})\ot\Delta_{l}
-\epsilon(\phi_{1}\sigma_{3}\phi_{2}^{\dag}+\phi_{2}\sigma_{3}\phi_{1}^{\dag})
\ot\Sigma_{l}^{'},&\nonumber\\
&C_{q}^{R}\;=\;M_{q}^{\dag}(1-\phi_{1}^{\dag}\phi_{1}-\phi_2{}^{\dag}\phi_{2}
-\phi_{1}^{\dag}\phi_{2}\sigma_{3}-\sigma_{3}\phi_{2}^{\dag}\phi_{2}
)M_{q},&\nonumber\\
&C_{l}^{R}\;=\;M_{l}^{\dag}(1-\phi_{1}^{\dag}\phi_{1}-\phi_{2}^{\dag}\phi_{2}
-\epsilon\phi_{1}^{\dag}\phi_{2}\sigma_{3}
-\epsilon\sigma_{3}\phi_{2}^{\dag}\phi_{1}
)M_{l}&  \nonumber
\eee
\end{itemize}
All the $\phi$'s are quaternions that parametrize the Higgs field and $\epsilon$ is an integer taking the value -1 (type 3 models) or +1 (type 4 models).
We use here  the notations:
\bb
\Sigma_{q}^{'}=\Sigma_{q}-
\frac{\t\lp x\Sigma_{q}\Delta_{q}\rp+\epsilon\t\lp y\Sigma_{l}\Delta_{l}\rp }
{\t\lp x\Delta_{q}^{2}\rp+\t\lp y\Delta_{l}^{2}\rp}\Delta_{q}&,&
\Sigma_{l}^{'}=\Sigma_{l}-
\frac{\t\lp x\Sigma_{q}\Delta_{q}\rp+\epsilon\t\lp y\Sigma_{l}\Delta_{l}\rp }
{\t\lp x\Delta_{q}^{2}\rp+\t\lp y\Delta_{l}^{2}\rp}\Delta_{l}\nonumber\\
\Sigma_{q}\;=\;\frac{1}{2}(M_{u}M_{u}^{*}+M_{d}M_{d}^{*})&,&
\Sigma_{l}\;=\;\frac{1}{2}M_{e}M_{e}^{*}\nonumber\\
\Delta_{q}^{'}\;=\;
\frac{\t\lp y\Delta_{l}^{2}\rp}{\t\lp x\Delta_{q}^{2}\rp+\t\lp y\Delta_{l}^{2}\rp}\Delta_{q}&,&
\Delta_{l}^{'}\;=\;
\frac{\t\lp x\Delta_{q}^{2}\rp}{\t\lp x\Delta_{q}^{2}\rp+\t\lp y\Delta_{l}^{2}\rp}\Delta_{l}.  \nonumber
\eee
Note that the axioms of NCG allow us to build models (for example take $\alpha=\beta=b$ and $\gamma=b'$) that yield {\it three} doublets of complex scalar fields. Requiring the existence of a {\it vector like} unbroken photon eliminates all possibilities with three doublets whose computation would be tedious.\\
Let us now estimate the mass of the new gauge boson.

\subsection{ Masses of the neutral gauge bosons}
In NCG, the bosonic part of the action is given by the square norm of the curvature of a gauge field on the noncommutative space time. It turns out to be (see \cite{SZ} for a detailed computation)
$$
S\lb X,\Phi\rb=\int_{\mm} Tr(z\tilde\rho(Y^{*})\wedge*\tilde\rho(Y))+Tr(zD\Phi^{*}\wedge*D\Phi)+*V(\Phi), 
$$
where $\tilde\rho$ is the representation of the Lie algebra of the gauge group $G$, $Y$ is the field strength of a genuine (Yang-Mills) gauge field $X$, $*$ is the Hodge star, $\wedge$ is the wedge product and $D\Phi=d\Phi+\lb\tilde\rho(X),\Phi\rb$ is the covariant derivative of $\Phi$. $V(\Phi)$ is no longer given by (\ref
{prelpot}) but is
\bb
&V(\Phi)=\t\lb\lp C-\rho(x)-\theta\rp^*\lp C-\rho(x)-\theta\rp\rb,&\nonumber
\eee
where $x\in\aa$ and $\theta\in J^2$ are determined by the condition that $C-\rho(x)-\theta$ be orthogonal to $\rho(\aa)\,+\,J^2$.  
To get the mass term for gauge bosons, we follow the usual approach of symmetry breaking in NCG and we first look for the minimum of $V(\Phi)$. The absolute minimum of $V(\Phi)$ is reached for $\Phi$ such that $C\in\rho(\aa)\,+\,J^2$ that is equivalent to $C=0$. Using equation (\ref{curv}), it is obvious that $C=0$ when $H=0$ that is when $\Phi=-iD$ up to gauge transformations. The converse also holds as one checks using the explicit expression of the curvature given in the previous section. Then, we replace $\Phi$ by $-iD$ in the expression of the covariant derivative of $\Phi$ to obtain the mass terms for gauge bosons. Notice that we do not compute the explicit form of the Higgs potential since we will not compute the scalar masses.\\
To proceed, we parametrize $X=X_{\mu}dx^{\mu}$  in the following manner:
\bb
&X_{\mu}=\frac{1}{2}\lp g_{2}\sigma_{a}A_{\mu}^{a},\> g_{1}\cos\theta B_{\mu}-g_{1^{'}}\sin\theta B_{\mu}^{'},\> g_{1}\sin\theta B_{\mu}+g_{1^{'}}\cos\theta B_{\mu}^{'},\>g_{3}\lambda^{a}C_{\mu}^{a}\rp&, \nonumber
\eee
where $\sigma_{a}$ and $\lambda_{a}$ denote the Pauli and Gell-Mann matrices normalized by $Tr(\lambda_{a}\lambda_{b})=2\delta_{ab}$. Due to the unimodularity condition, the two $u(1)$ summands of the Lie algebra $\gg$ are no longer orthogonal so we have to introduce a mixing angle $\theta$. The coupling constants $g_{2}$, $g_{1}$, $g_1'$ and $g_{3}$ and the mixing angle $\theta$  are determined such that the spin 1 part of our action becomes the usual action of a Yang-Mills theory:
\begin{displaymath}
\int_{\mm}Tr(z\rho(Y)^{*}\wedge *\rho(Y))=\int_{\mm} \lp\frac{1}{4}F_{\mu\nu}^{a}F^{a\mu\nu}+\frac{1}{4}G_{\mu\nu}G^{\mu\nu}+\frac{1}{4}G_{\mu\nu}^{'}G^{'\mu\nu}+\frac{1}{4}H_{\mu\nu}^{a}H^{a\mu\nu}\rp d^{4}x,
\end{displaymath}
where $F_{\mu\nu}^{a}$, $G_{\mu\nu}$, $G^{'}_{\mu\nu}$ and $H_{\mu\nu}^{a}$ are the usual Yang-Mills curvatures associated to the gauge fields $A_{\mu}^{a}$, $B_{\mu}$,$B_{\mu}^{'}$ and $C_{\mu}^{'}$. Then, after replacing $\Phi$ by $-iD$ in the square norm of the covariant derivative of the scalar field, we obtain the mass term for the gauge bosons as a quadratic form of $A_{\mu}^a$, $B_{\mu}$ and $B_{\mu}'$. With respect to this quadratic form, the charged gauge boson $A_{\mu}^1$ and $A_{\mu}^2$ are orthogonal to the neutral gauge bosons $A_{\mu}^3$, $B_{\mu}$ and $B_{\mu}'$ and in the vector space spanned by the three neutral gauge bosons, this quadratic form is:
\bb
&\pp{A_{\mu}^{3}&B_{\mu}&B^{'}_{\mu}}
\pp{M^{2}_{AA}&M^{2}_{AB}&M^{2}_{AB^{'}}\cr
M^{2}_{AB}&M^{2}_{BB}&0\cr
M^{2}_{AB^{'}}&0&M^{2}_{B^{'}B^{'}}}
\pp{A^{3\mu}\cr B^{\mu}\cr B^{'\mu}},&\nonumber
\eee
where
\bb
&M^{2}_{AA}=\frac{1}{2}g_{2}^{2}\lp Tr(xM_{u}M_{u}^{*})
+Tr(xM_{d}M_{d}^{*})
+Tr(yM_{e}M_{e}^{*})\rp,& \nonumber
\eee
\bb
&M^{2}_{BB}=
\frac{1}{2}g_{1}^{2}\lp 
Tr(xM_{u}M_{u}^{*})\lp y_{u}\cos^{2}\theta +{y'}_{u}\sin^{2}\theta\rp\rp&\nonumber\\
&+Tr(xM_{d}M_{d}^{*}))\lp y_{d}\cos^{2}\theta +{y'}_{d}\sin^{2}\theta\rp
+Tr(yM_{e}M_{e}^{*}))\lp y_{e}\cos^{2}\theta +{y'}_{e}\sin^{2}\theta\rp
,& \nonumber
\eee
\bb
&M^{2}_{B'B'}=\frac{1}{2}g_{1'}^{2}\lp 
Tr(xM_{u}M_{u}^{*})\lp y'_{u}\cos^{2}\theta +y_{u}\sin^{2}\theta\rp\rp&\nonumber\\
&+Tr(xM_{d}M_{d}^{*})\lp y'_{d}\cos^{2}\theta+y_{d}\sin^{2}\theta\rp
+Tr(yM_{e}M_{e}^{*})\lp y'_{e}\cos^{2}\theta +y_{e}\sin^{2}\theta\rp
,& \nonumber
\eee
and the coupling constants $g_{2}$, $g_{1}$ and $g_{1}'$ and the mixing angle $\theta$ are given by
\bb
&g_{2}^{-2}=Nx+Tr(y),& \nonumber
\eee
\bb
&g_{1}^{-2}=
 \frac{1}{2}\cos^{2}\theta\lp y_{u}^{2}Nx+y_{d}^{2}Nx+y_{e}^{2}Tr(y)+4N\tilde{x}u^{2}+3Tr(\tilde{y})\rp&\nonumber\\
&+\frac{1}{2}\sin^{2}\theta\lp
{y'}_{u}^{2}Nx+{y'}_{d}^{2}Nx+{y'}_{e}^{2}Tr(y)+4N\tilde{x}'u^{2}\rp
-4N\tilde{x}uu'\cos\theta\sin\theta,& \nonumber
\eee
\bb
&g_{1'}^{-2}=
\frac{1}{2}\sin^{2}\theta\lp y_{u}^{2}Nx+y_{d}^{2}Nx+y_{e}^{2}Tr(y)+4N\tilde{x}u^{2}+3Tr(\tilde{y})\rp&\nonumber\\
&+\frac{1}{2}\cos^{2}\theta\lp
{y'}_{u}^{2}Nx+{y'}_{d}^{2}Nx+{y'}_{e}^{2}Tr(y)+4N\tilde{x}u'^{2}\rp
+4N\tilde{x}uu'\cos\theta\sin\theta,& \nonumber
\eee
$$
\tan 2\theta=
\frac{8N\tilde{x}uu'}{2Nx\lp y_{u}^{2}+y_{d}^{2}-1\rp +Tr(y)\lp 2y_{e}^{2}-1\rp+4N\tilde{x}\lp u^{2}-u^{'2}\rp+6Tr(\tilde{y})}\, .
$$
For our purpose we do not need the explicit form of $M^{2}_{AB}$ and $M^{2}_{AB^{'}}$. To end up we should diagonalize this matrix in an orthonormal basis (for the scalar product defined by $z$) to obtain as eigenstates the photon and the two massive neutral bosons. We simply take the trace of the mass matrix and we get:
\bb
&\frac{1}{2}m_{Z}^{2}+\frac{1}{2}m_{Z^{'}}^{2}=
M_{AA}^{2}+M_{BB}^{2}+M_{B^{'}B^{'}}^{2}&\label{trace}.
\eee 
Although the right hand side depends on the noncommutative gauge coupling $z$, we can derive an inequality between the masses of the neutral gauge bosons and the masses of the fermions. To proceed, we first notice that $\tilde x$ and $\tilde y$ only appear in the denominators of the gauge couplings as positive quantities so that the following inequalities hold:
$$
g_{1}^2<\frac{2}
{\cos^{2}\theta\lp\alpha^2Nx+\beta^2Nx+\gamma^2\t(y)\rp
+\sin^2\theta\lp\alpha'^2Nx+\beta'^2Nx+\gamma'^2\t(y)\rp},\nonumber
$$
$$
g_{1}'^2<\frac{2}
{\sin^{2}\theta\lp\alpha^2Nx+\beta^2Nx+\gamma^2\t(y)\rp
+\cos^2\theta\lp\alpha'^2Nx+\beta'^2Nx+\gamma'^2\t(y)\rp}.\nonumber
$$
Replacing these inequalities in ($\ref{trace}$), we obtain an inequality whose right hand side is a sum of quotients of linear functions of $x$ and $y$. We use the following inequality involving the strictly positive numbers $a_1$, $a_2$, $a_!'$, $a_2'$, $x_!$ and $x_2$:
$$
\frac{a_1x_1+a_2x_2}{a_1'x_1+a_2'x_2}<\frac{a_1}{a_1'}+\frac{a_2}{a_2'},
$$
to eliminate the parameters $x$ and $y$ of the noncommutative gauge coupling and the mixing terms $\cos^2\theta$ and $\sin^2\theta$. The result is independent of the values of the generalized hypercharges:
$$
\frac{1}{2}m_{Z}^{2}+\frac{1}{2}m_{Z'}^{2}<\frac{3}{2N}\lp \t(M_uM_u^*)+\t(M_dM_d^*)\rp+3\t(M_eM_e^*)^2.
$$
If we neglect all fermions' masses but the top mass, we get
\begin{equation}
m_{Z}^{2}+m_{Z'}^{2}<m_{t}^{2}\nonumber
\end{equation}

This means that it is impossible to construct
a new gauge boson with the mass bigger than 
170 Gev. However, this constraint can be removed by adding a fourth family of fermions that contains at least one heavy particle.

\section{Conclusion and outlook}

In this work, we have studied all $U(1)$ extensions of the SM derived from NCG with the algebra $\hhh\op\cc\op\cc\op M_{3}(\cc)$ and without adding new fermions. It appears that the extra $U(1)$ gauge boson is always anomalous (even in the more general YMH framework) and that its mass cannot exceed the mass of the heaviest fermion. Therefore, if we want to constuct a heavy $Z'$ gauge boson in NCG, we should add at least one heavy fermion to the SM. This result has to be compared with \cite{uni} and \cite{PS} who have studied some extensions of the SM in NCG. To us, this is an indication of the distinguished character of the SM in NCG in the sense that mild extensions of the SM are not available in NCG.\\
We end this paper with the conjecture that, in a general NCG theory, the masses of the spontaneously broken generators lie between the lightest and the heaviest fermions.\\

\vskip 0.5truecm
\noindent
{\bf\Large Aknowledgements}\\
\noindent
It is a pleasure for us to thank B. Iochum and T. Sch\"ucker for their helpful advices.
\vskip 0.5truecm


 \end{document}